\documentclass[twoside,twocolumn,english,aps,prl,showpacs]{revtex4}
\usepackage[T1]{fontenc}
\usepackage[latin1]{inputenc}
\usepackage{graphicx}

\makeatletter

\providecommand{\LyX}{L\kern-.1667em\lower.25em\hbox{Y}\kern-.125emX\@}

\usepackage{babel}
\makeatother
\begin{document}

\newcommand{\pd}{\hat{\psi }^{\dagger }}

\newcommand{\ps}{\hat{\psi }}
 
\newcommand{\ph}{\hat{\phi }}

\newcommand{\kt}[1]{\left|#1\right\rangle }

\title{Influence of dephasing on shot noise in an electronic Mach-Zehnder
interferometer}

\author{Florian Marquardt and C. Bruder}

\affiliation{Departement Physik und Astronomie, Universität Basel, Klingelbergstr.
82, 4056 Basel, Switzerland}

\date{17.6.2003}

\begin{abstract}
We present a general analysis of shot noise in an electronic Mach-Zehnder
interferometer, of the type investigated experimentally {[}Yang Ji
\emph{et al.}, Nature 422, 415 (2003){]}, under the influence of dephasing
produced by fluctuations of a classical field. We show how the usual
partition noise expression $\mathcal{T}(1-\mathcal{T})$ is modified
by dephasing, depending on the power spectrum of the environmental
fluctuations. 
\end{abstract}


\maketitle

\newcommand{\p}{\varphi }

A large part of mesoscopic physics is concerned with quantum interference
effects in micrometer-size electronic circuits. Therefore, it is important
to understand how interference is suppressed by the action of a fluctuating
environment (such as phonons or other electrons), a phenomenon known
as dephasing (or decoherence). In recent years, many experimental
studies have been performed to learn more about the mechanisms of
dephasing and its dependence on parameters such as temperature \cite{key-11,key-15,key-13,hansen,key-12,key-25,key-28}. 

A delicate issue \cite{hansen} in the analysis of interference effects
other than weak localization is the fact that the {}``visibility''
of the interference pattern can also be diminished by phase averaging:
This takes place when electrons with a spread of wavelengths (determined
by voltage or temperature) contribute to the current, or when some
experimental parameter (such as path length) fluctuates slowly from
run to run. Recently, a remarkable interference experiment has been
performed using a Mach-Zehnder setup fabricated from the edge channels
of a two-dimensional electron gas in the integer quantum hall effect
regime \cite{heiblum}. Besides measuring the current as a function
of the phase difference between the paths, the authors also measured
the shot noise to distinguish between phase averaging and {}``real''
dephasing. The basic idea is that both phenomena suppress the interference
term in the current, but they may affect differently the partition
noise which is nonlinear in the transmission probability. The idea
of using shot noise to learn more about dephasing is promising, connecting
two fundamental issues in mesoscopic physics. 

\begin{figure}
\includegraphics[  width=3in,
  keepaspectratio]{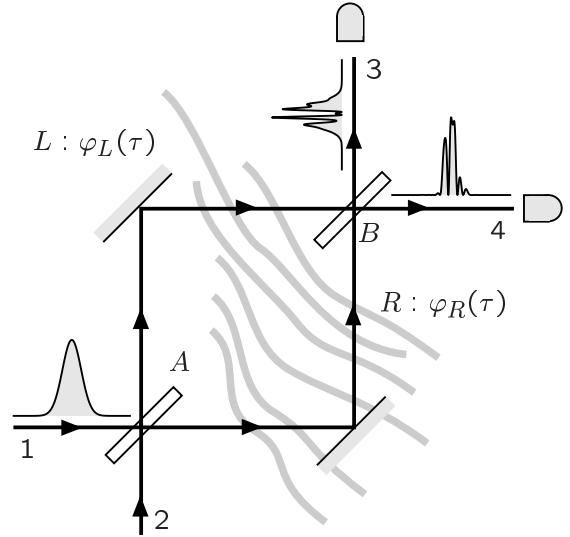}

\caption{\label{cap:fig1}The Mach-Zehnder interferometer setup analyzed in
the text. In the case shown here, the fluctuations of the environment
are fast compared with the temporal extent of the wave packet (determined
by temperature or voltage, see text). The probability density of the
incoming wave packet and its two outgoing parts is shown.}
\end{figure}

Most theoretical works on dephasing in mesoscopic interference setups
are concerned with its influence on the average current only (see
Refs.~\onlinecite{aleiner,levinson,seelig,marquardt,marquardtDDD,seelig2}
and references therein), although there have been studies of shot
noise in this context \cite{qheshotnoise}. In this paper, we present
the first analysis of shot noise for an electronic one-channel Mach-Zehnder
interferometer under the influence of dephasing. We will consider
dephasing produced by fluctuations of a classical potential, which
describe either true nonequilibrium radiation impinging on the system
or the thermal part of the environmental noise. This approach has
been employed quite often in the past \cite{classnoise,seelig}, is
exact in the first case, and should be a reliable approximation for
$T\gg eV$ in the second case. In particular, we are interested in
the influence of the power spectrum of the environmental fluctuations
on the shot-noise result, a question that goes beyond the phenomenological
dephasing terminal model \cite{buettinelastic,key-9,deJongBeen,vanlangen,qheshotnoise}. 

\emph{Model and general results. -} We consider non-interacting, spin-polarized
electrons. By solving the Heisenberg equation of motion for the electron
field $\hat{\Psi }$ moving at constant velocity $v_{F}$ (linearized
dispersion relation) and under the action of a fluctuating potential
$V(x,t)$ (without backscattering), we obtain 

\begin{equation}
\hat{\Psi }(x,\tau )=\int \frac{dk}{\sqrt{2\pi }}e^{-i\epsilon _{k}\tau }\sum _{\alpha =1}^{3}t_{\alpha }(k,\tau )\hat{a}_{\alpha }(k)e^{s_{\alpha }ik_{F}x}\label{eq:psi}\end{equation}
for the electron operator at the output port $3$. We have $t_{3}=1$,
$s_{1,2}=1,\, s_{3}=-1$, the reservoir operators obey $\left\langle \hat{a}_{\alpha }^{\dagger }(k)\hat{a}_{\beta }(k')\right\rangle =\delta _{\alpha \beta }\delta (k-k')f_{\alpha }(k)$,
and the integration is over $k>0$ only. The amplitudes $t_{1},\, t_{2}$
for an electron to go from terminal $1$ or $2$ to the output terminal
$3$ are time-dependent:

\begin{eqnarray}
t_{1}(k,\tau ) & = & t_{A}t_{B}e^{i\varphi _{R}(\tau )}+r_{A}r_{B}e^{i\varphi _{L}(\tau )}e^{i(\phi +k\delta x)}\label{eq:t1}\\
t_{2}(k,\tau ) & = & t_{A}r_{B}e^{i\varphi _{L}(\tau )}e^{i(\phi +k\delta x)}+r_{A}t_{B}e^{i\varphi _{R}(\tau )}\label{eq:t2}
\end{eqnarray}
Here $t_{A/B}$ and $r_{A/B}$ are energy-independent transmission
and reflection amplitudes at the two beamsplitters ($t_{j}^{*}r_{j}=-t_{j}r_{j}^{*}$),
$\delta x$ is a possible path-length difference, and $\phi $ the
Aharonov-Bohm phase due to the flux through the interferometer. The
electron accumulates fluctuating phases while moving along the left
or right arm: $\varphi _{L,R}(\tau )=-\int _{-\tau _{L,R}}^{0}dt'\, V(x_{L,R}(t'),\tau +t')$,
where $\tau $ is the time when the electron leaves the second beamsplitter
after traveling for a time $\tau _{L,R}$ along $x_{L,R}$. Note that
in our model the total traversal times $\tau _{L,R}$ enter only at
this point.

The output current following from (\ref{eq:psi}) has to be averaged
over the fluctuating phases, i.e. it depends on phase-averaged transmission
probabilities $T_{1}=\left|t_{1}\right|^{2}$ and $T_{2}=1-T_{1}$: 

\begin{equation}
\left\langle T_{1}\right\rangle _{\varphi }=T_{A}T_{B}+R_{A}R_{B}+2z\left(r_{A}r_{B}\right)^{*}t_{A}t_{B}\cos (\phi +k\delta x),\label{eq:prob}\end{equation}
The interference term is suppressed by $z\equiv \left\langle e^{i\delta \varphi }\right\rangle _{\varphi }$
(with $\delta \varphi =\varphi _{L}-\varphi _{R}$), which is $\exp (-\left\langle \delta \varphi ^{2}\right\rangle /2)$
for Gaussian $\delta \varphi $. This decreases the visibility of
the interference pattern observed in $I(\phi )$. However, such a
suppression can also be brought about by the $k$-integration, if
$\delta x\neq 0$ (thermal smearing).

Our main goal is to calculate the shot noise power $S$ at zero frequency.
It can be split into two parts:

\begin{eqnarray}
S=\int d\tau \, \left\langle \left\langle \hat{I}(\tau )\hat{I}(0)\right\rangle \right\rangle _{\varphi }-\left\langle \left\langle \hat{I}(0)\right\rangle \right\rangle _{\varphi }^{2}= &  & \nonumber \\
\int d\tau \, \left\langle \left\langle \hat{I}(\tau )\right\rangle \, \left\langle \hat{I}(0)\right\rangle \right\rangle _{\varphi }-\left\langle \left\langle \hat{I}(0)\right\rangle \right\rangle _{\varphi }^{2}+ &  & \nonumber \\
\int d\tau \, \left\langle \left\langle \hat{I}(\tau )\hat{I}(0)\right\rangle -\left\langle \hat{I}(\tau )\right\rangle \left\langle \hat{I}(0)\right\rangle \right\rangle _{\varphi } &  & 
\end{eqnarray}
The first integral on the r.h.s. describes shot noise due to the temporal
fluctuations of the conductance, i.e. fluctuations of a classical
current $I(\tau )=\left\langle \hat{I}(\tau )\right\rangle $ depending
on time-dependent transmission probabilities. We denote its noise
power as $S_{\textrm{cl}}$. It rises quadratically with the total
current, as is known from $1/f$-noise in mesoscopic conductors \cite{kogan}. 

The second integral is evaluated by inserting (\ref{eq:psi}) and
applying Wick's theorem:

\begin{widetext}

\begin{equation}
\left\langle \left\langle \hat{I}(\tau )\hat{I}(0)\right\rangle -\left\langle \hat{I}(\tau )\right\rangle \left\langle \hat{I}(0)\right\rangle \right\rangle _{\varphi }=\left(\frac{ev_{F}}{2\pi }\right)^{2}\int dkdk'\, \sum _{\alpha ,\beta =1,2,3}f_{\alpha }(k)(1-f_{\beta }(k'))\, K_{\alpha \beta }(\tau )e^{i(\epsilon _{k'}-\epsilon _{k})\tau }.\label{eq:wideformula}\end{equation}
\end{widetext}Here $K_{\alpha \beta }$ is a correlator of four transmission
amplitudes: We have $K_{33}=1$, $K_{3\alpha }=K_{\alpha 3}=0$, and

\begin{equation}
K_{\alpha \beta }(\tau )\equiv \left\langle t_{\alpha }^{*}(k,\tau )t_{\beta }(k',\tau )t_{\alpha }(k,0)t_{\beta }(k',0)^{*}\right\rangle _{\varphi }\, ,\label{eq:AmplCorr}\end{equation}
for $\alpha ,\beta =1,2$. 

In order to obtain two limiting forms of this expression, we note
that the $\tau $-range of the oscillating exponential factor under
the integral in (\ref{eq:wideformula}) is determined by the Fermi
functions, i.e. by voltage and temperature. This has to be compared
with the correlation time $\tau _{c}$ of the environment (the typical
decay time of the phase correlator $\left\langle \delta \varphi (\tau )\delta \varphi (0)\right\rangle $).
For $eV\tau _{c}\ll 1$ and $T\tau _{c}\ll 1$ ({}``fast environment''),
the major contribution of the integration comes from $|\tau |\gg \tau _{c}$,
where $K_{\alpha \beta }$ factorizes into

\begin{equation}
K_{\alpha \beta }(\tau )\approx K_{\alpha \beta }(\infty )\equiv \left|\left\langle t_{\alpha }^{*}(k,0)t_{\beta }(k',0)\right\rangle _{\varphi }\right|^{2}\, .\label{eq:Klimit}\end{equation}

This yields the noise power

\begin{equation}
\frac{S_{\textrm{fast}}}{e^{2}v_{F}/2\pi }=\int dk\, \sum _{\alpha ,\beta =1,2}f_{\alpha }(1-f_{\beta })\left|\left\langle t_{\alpha }^{*}t_{\beta }\right\rangle _{\varphi }\right|^{2}\, ,\label{eq:Sfast}\end{equation}
where we have set $f_{\alpha ,\beta }=f_{\alpha ,\beta }(k)$ and
$t_{\alpha ,\beta }=t_{\alpha ,\beta }(k,0)$. We conclude that the
shot noise for a {}``fast'' environment is \emph{not} given by a
simple expression of the form $\left\langle \mathcal{T}\right\rangle _{\varphi }(1-\left\langle \mathcal{T}\right\rangle _{\varphi })$,
since we have

\begin{equation}
\left|\left\langle t_{1}^{*}t_{2}\right\rangle _{\varphi }\right|^{2}-\left\langle T_{1}\right\rangle _{\varphi }(1-\left\langle T_{1}\right\rangle _{\varphi })=(z^{2}-1)R_{B}T_{B}\, .\end{equation}

The remainder of the noise power from Eq. (\ref{eq:wideformula})
(with $K_{\alpha \beta }(\tau )-K_{\alpha \beta }(\infty )$ inserted
in Eq. (\ref{eq:wideformula})) will be denoted $S_{\textrm{fluct}}$.
It yields a contribution to the Nyquist noise $S_{V=0}$, but apart
from that it becomes important only at larger $V,\, T$. With this
definition, the full noise power can be written as

\begin{equation}
S=S_{\textrm{fast}}+S_{\textrm{fluct}}+S_{\textrm{cl}}\, ,\label{eq:S}\end{equation}
which is valid over the entire parameter space.

In the other limiting case, when the $\tau $-integration is dominated
by $|\tau |\ll \tau _{c}$ ({}``slow environment''), we can use
$K_{\alpha \beta }(\tau )\approx K_{\alpha \beta }(0)$, which yields

\begin{equation}
\frac{S_{\textrm{slow}}}{e^{2}v_{F}/2\pi }=\int dk\, \left\langle (f_{1}T_{1}+f_{2}T_{2})(1-(f_{1}T_{1}+f_{2}T_{2}))\right\rangle _{\varphi }\, ,\label{eq:Sslow}\end{equation}
i.e. the phase-average of the usual shot noise expression (at $T=0$
the integrand is $\left\langle T_{1}(1-T_{1})\right\rangle _{\varphi }$). 

\begin{figure}
\includegraphics[  width=3in,
  keepaspectratio]{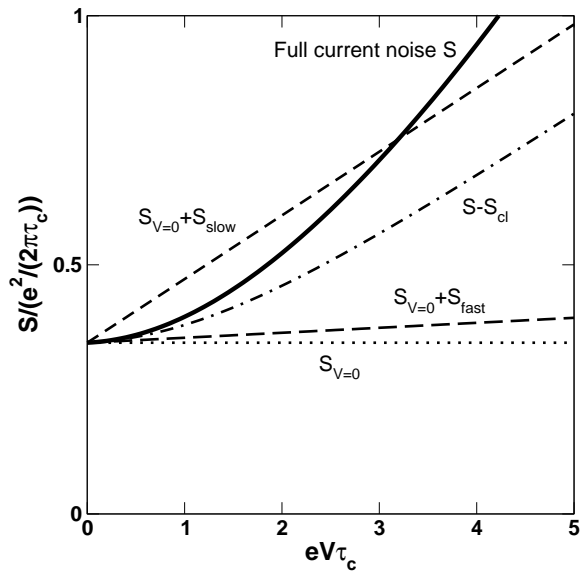}

\caption{\label{cap:fignoiseV}Typical behaviour of the full current noise
$S$ as a function of $eV\tau _{c}$. At higher voltages, the dependence
on $V$ is quadratic, due to $S_{\textrm{cl}}$. When $S_{\textrm{cl}}$
is subtracted, the slope at large $eV\tau _{c}$ is determined by
$S_{\textrm{slow}}$ (i.e. $\left\langle T_{1}(1-T_{1})\right\rangle _{\varphi }$),
while that at low voltages is always determined by $S_{\textrm{fast}}$
(i.e. $\left|\left\langle t_{1}^{*}t_{2}\right\rangle _{\varphi }\right|^{2}$).
Parameters: $T=0$, $\delta x=0$, $\phi =0$, $T_{A}=1/2$, $z=1/e$,
$T_{B}=0.4$.}
\end{figure}
\emph{Discussion. -} We are able to evaluate the phase-averages if
the potential $V(x,t)$ (and therefore $\delta \varphi $) is assumed
to be a Gaussian random field of zero mean. In the following, we present
explicit expressions for the case $T=0$, $\delta xeV/v_{F}\ll 1$,
where the visibility is decreased purely by dephasing. We can express
the results by the following Fourier transforms that depend on the
power spectrum of the fluctuations and are nonperturbative in the
fluctuating field ($\lambda =\pm $):

\begin{eqnarray}
\hat{g}_{\lambda }(\omega ) & \equiv  & \int d\tau \, e^{i\omega \tau }[e^{\lambda \left\langle \delta \varphi (\tau )\delta \varphi (0)\right\rangle }-1],\\
I_{\lambda }(V) & \equiv  & \int _{0}^{eV}d\omega \, (1-\frac{\omega }{eV})\hat{g}_{\lambda }(\omega )\label{eq:Idef}
\end{eqnarray}
The shot noise becomes ($\tilde{\phi }=\phi +k_{F}\delta x$):

\begin{eqnarray}
\frac{S-S_{V=0}}{e^{3}V/2\pi } & = & \frac{eV}{\pi }\, z^{2}R_{A}R_{B}T_{A}T_{B}(\cos (2\tilde{\phi })\hat{g}_{-}(0)+\hat{g}_{+}(0))\nonumber \\
 &  & +\left|\left\langle t_{1}^{*}t_{2}\right\rangle _{\varphi }\right|^{2}\nonumber \\
 &  & +\frac{1}{\pi }\, z^{2}R_{B}T_{B}\left\{ -2\cos (2\tilde{\phi })R_{A}T_{A}\, I_{-}(V)\right.\nonumber \\
 &  & \left.+(R_{A}^{2}+T_{A}^{2})\, I_{+}(V)\right\} \label{eq:shotnoiseT0}
\end{eqnarray}
The first line corresponds to $S_{\textrm{cl}}$, the second to $S_{\textrm{fast}}$,
and the rest to $S_{\textrm{fluct}}-S_{V=0}$. At $V\rightarrow 0$,
the integrals $I_{\pm }(V)$ vanish and $S_{\textrm{fast}}$ dominates.
At large $eV\tau _{c}\gg 1$ we can use the sum-rule $I_{\lambda }(V)\rightarrow \pi \left[z^{-2\lambda }-1\right]$
and find the last three lines to combine to $\left\langle T_{1}(1-T_{1})\right\rangle _{\varphi }$,
i.e. $S_{\textrm{slow}}$. The Nyquist noise is $\phi $-independent: 

\begin{equation}
S_{V=0}=\frac{e^{2}}{2\pi ^{2}}z^{2}R_{B}T_{B}\int _{0}^{\infty }d\omega \, \omega \hat{g}_{+}(\omega )\, .\label{eq:nyquist}\end{equation}

\begin{figure}
\includegraphics[  width=3in]{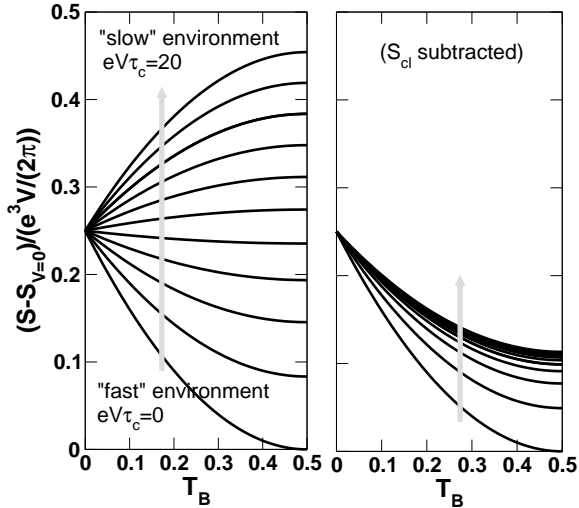}

\caption{\label{figSNze4}Normalized shot noise $(S-S_{V=0})/(e^{3}V/2\pi )$
as a function of the transmission $T_{B}$ of the second beamsplitter
for small visibility ($z=1/e$). The different curves show the succession
from a {}``fast'' environment to a {}``slow'' one (bottom to top
: $eV\tau _{c}=0,\, 2,\, 4,\, \ldots ,\, 20$). In the right panel,
the contribution from $S_{cl}$ (first line of (\ref{eq:shotnoiseT0}))
has been dropped, to demonstrate the convergence against the result
for a slow bath, $\left\langle T_{1}(1-T_{1})\right\rangle _{\varphi }$
(topmost curve). Other parameters as in Fig.~\ref{cap:fignoiseV}.}
\end{figure}

The results are illustrated in Figs. \ref{cap:fignoiseV} and \ref{figSNze4},
where the evolution of $S$ with increasing voltage $V$ shows the
cross-over between a {}``fast'' and a {}``slow'' environment.
Although $S_{\textrm{fast}}$ can become zero, the total current noise
$S$ does not vanish, due to the Nyquist contribution. The plots have
been produced by assuming 50\% transparency of the first beamsplitter
($T_{A}=1/2$) and using a simple Gaussian form for the phase correlator:

\begin{equation}
\left\langle \delta \varphi (\tau )\delta \varphi (0)\right\rangle =\left\langle \delta \varphi ^{2}\right\rangle e^{-(\tau /\tau _{c})^{2}}\, .\end{equation}
An application of the general theory presented here to specific situations
includes the calculation of the phase-correlator, starting from the
correlator describing the potential fluctuations $V(x,t)$ (cf. \cite{seelig}
for an example).

If shot noise is to be used as a tool to distinguish phase averaging
from 'genuine dephasing', it is important to inquire about the form
of shot noise for a situation in which both effects are present. In
particular, for a large path-length difference $\delta x\gg v_{F}/eV$
(relevant for the experiment of Ref.~\cite{heiblum}), or $\delta x\gg v_{F}/T$,
the interference term is already suppressed completely due to wavelength
averaging. Then $S_{\textrm{cl}}$ vanishes, since $\left\langle \hat{I}(\tau )\right\rangle $
is independent of the fluctuating phase. In addition, we find (at
$T=0$):

\begin{eqnarray}
\frac{S-S_{V=0}}{e^{3}V/2\pi } & = & T_{A}R_{A}(T_{B}-R_{B})^{2}+\label{eq:largedx1}\nonumber \\
 &  & z^{2}T_{B}R_{B}(T_{A}^{2}+R_{A}^{2})[1+\frac{I_{+}(V)}{\pi }]\, .\label{eq:largedx}
\end{eqnarray}
 For a {}``fast'' environment, we have $I_{+}(V)\rightarrow 0$,
such that Eq. (\ref{eq:largedx}) becomes $\left[(T_{B}-R_{B})^{2}+2z^{2}R_{B}T_{B}\right]/4$,
which turns into $(T_{B}-R_{B})^{2}/4$ for $z\rightarrow 0$. This
could be distinguished from the $k$-averaging result, but it corresponds
to the case of large energy transfers from and to the environment
(as opposed to {}``pure dephasing''). On the other hand, in the
limit of large voltages ({}``slow environment'', $eV\tau _{c}\gg 1$),
we have $I_{+}(V)\rightarrow \pi \left[z^{-2}-1\right]$, and Eq.
(\ref{eq:largedx}) turns into $(T_{B}^{2}+R_{B}^{2})/4$, which is
equal to the result obtained for pure $k$-averaging (without any
additional dephasing).

We have pointed out already that neither $S_{\textrm{fast}}$ nor
$S_{\textrm{slow}}$ lead to the simple result $\left\langle T_{1}\right\rangle _{\varphi }(1-\left\langle T_{1}\right\rangle _{\varphi })$.
However, this form does indeed apply if we consider injecting only
a {}``narrow beam'' of electrons into terminal $1$ (i.e. $f_{\alpha }(k)=0$
except for $f_{1}(k)=1$ within $[k_{F},k_{F}+eV/v_{F}]$), which
is not equivalent to the previous situation regarding shot noise (cf.
\cite{gavish} in this respect). We get for $eV\tau _{c}\ll 1$:

\begin{equation}
S-S_{\textrm{cl}}=\frac{e^{3}V}{2\pi }\left\langle T_{1}\right\rangle _{\varphi }(1-\left\langle T_{1}\right\rangle _{\varphi })\, ,\label{eq:snbeam}\end{equation}
while the case $eV\tau _{c}\gg 1$ is described by $S_{\textrm{slow}}$. 

\emph{Comparison with other models.} - Finally, we compare our results
in the fully incoherent limit ($z=0$) with two other models, namely
the phenomenological dephasing terminal \cite{deJongBeen,vanlangen,qheshotnoise},
and a simple model of a stream of regularly injected electrons \cite{key-24}
that reach the output port with a probability calculated according
to classical rules. We focus on $T=0$ and the case $T_{A}=1/2$.
At small path-length difference $eV\delta x/v_{F}\ll 1$ (no $k$-averaging),
we obtain $\left\langle T_{1}\right\rangle _{\varphi }(1-\left\langle T_{1}\right\rangle _{\varphi })=1/4$
both for the classical model and the narrow beam of electrons, $(T_{B}-R_{B})^{2}/4$
for our shot-noise expression in the {}``fast'' case, and $(T_{B}^{2}+R_{B}^{2})/4$
both for the {}``slow'' case and from the dephasing terminal \cite{note}.
In the opposite limit of large $\delta x$ only the result for the
classical model changes, coinciding with the {}``slow'' case $(T_{B}^{2}+R_{B}^{2})/4$,
which is also obtained without any dephasing. Therefore, in this regime
a shot noise measurement most likely will not be able to reveal the
additional presence of dephasing. 

In conclusion, we have derived shot-noise expressions for an experimentally
relevant model of an electronic Mach-Zehnder interferometer under
the influence of dephasing by a classical fluctuating potential. The
dependence of the shot noise on the power spectrum of the fluctuations
has been analyzed, demonstrating a cross-over between two regimes.
We have pointed out that a shot-noise measurement cannot reveal the
presence of dephasing on top of thermal averaging, for environmental
fluctuations slower than the inverse voltage or temperature. The present
theory may be applied to other interferometer geometries as well,
even in the presence of backscattering at junctions.

We thank M. Heiblum for useful comments and for sending us a preprint
of Ref.~\onlinecite{heiblum}. Useful discussions with W. Belzig,
C. Egues, M. Büttiker and E. Sukhorukov are gratefully acknowledged.
This work is supported by the Swiss NSF and the NCCR nanoscience.


\begin{thebibliography}{10}
\bibitem{key-11}P. Mohanty, E.~M.~Q. Jariwala, and R.~A. Webb, Phys. Rev. Lett.
\textbf{78}, 3366 (1997). 
\bibitem{key-15}E. Buks, R. Schuster, M. Heiblum, D. Mahalu, and H. Shtrikman, Nature
\textbf{391}, 871 (1998).
\bibitem{key-13}D. P. Pivin, A. Andresen, J. P. Bird, and D. K. Ferry, Phys. Rev.
Lett. \textbf{82}, 4687 (1999).
\bibitem{hansen}A. E. Hansen, A. Kristensen, S. Pedersen, C. B. Sørensen, and P. E.
Lindelof, Phys. Rev. B \textbf{64}, 045327 (2001).
\bibitem{key-12}D. Natelson, R. L. Willett, K. W. West, and L. N. Pfeiffer, Phys.
Rev. Lett. \textbf{86}, 1821 (2001).
\bibitem{key-25}K. Kobayashi, H. Aikawa, S. Katsumoto, and Y. Iye, J. Phys. Soc. Jpn.
\textbf{71}, L2094 (2002).
\bibitem{key-28}F. Pierre and N. O. Birge, Phys. Rev. Lett. \textbf{89}, 206804 (2002).
\bibitem{heiblum}Y. Ji, Y. Chung, D. Sprinzak, M. Heiblum, D. Mahalu, and H. Shtrikman,
Nature \textbf{422}, 415 (2003).
\bibitem{aleiner}I. L. Aleiner, N. S. Wingreen, and Y. Meir, Phys. Rev. Lett. \textbf{79},
3740 (1997).
\bibitem{levinson}Y. Levinson, Europhys. Lett. \textbf{39}, 299 (1997).
\bibitem{seelig}G. Seelig and M. Büttiker, Phys. Rev. B \textbf{64}, 245313 (2001).
\bibitem{marquardt}F. Marquardt and C. Bruder, Phys. Rev. B \textbf{65}, 125315 (2002). 
\bibitem{marquardtDDD}F. Marquardt and C. Bruder, cond-mat/0303397 (2003).
\bibitem{seelig2}G. Seelig, S. Pilgram, A. N. Jordan, and M. Büttiker, cond-mat/0304022
(2003).
\bibitem{qheshotnoise}C. Texier and M. Büttiker, Phys. Rev. B \textbf{62}, 7454 (2000).
\bibitem{classnoise}B.~L. Altshuler, A.~G. Aronov, and D.~E. Khmelnitsky, J. Phys.
C Solid State \textbf{15}, 7367 (1982); S. Chakravarty and A. Schmid,
Phys. Rep. \textbf{140}, 195 (1986). A. Stern, Y. Aharonov, and Y.
Imry, Phys. Rev. A \textbf{41}, 3436 (1990). 
\bibitem{buettinelastic}M. Büttiker, Phys. Rev. B \textbf{33}, 3020 (1986).
\bibitem{key-9}M. J. M. de Jong and C. W. J. Beenakker, in \emph{Mesoscopic Electron
Transport}, ed. by L. P. Kouwenhoven, L. L. Sohn, and G. Schön, NATO
ASI Series Vol. 345 (Kluwer Academic, Dordrecht, 1997).
\bibitem{deJongBeen}M.J.M. de Jong and C. W. J. Beenakker, Physica A \textbf{230}, 219
(1996).
\bibitem{vanlangen}S. A. van Langen and M. Büttiker, Phys. Rev. B \textbf{56}, R1680
(1997).
\bibitem{kogan}Sh. Kogan: {}``\emph{Electronic noise and fluctuations in solids}''
(Cambridge Univ. Press, Cambridge 1996).
\bibitem{gavish}U. Gavish, Y. Levinson, and Y. Imry, Phys. Rev. Lett. \textbf{87},
216807 (2001). 
\bibitem{key-24}Ya. M. Blanter and M. Büttiker, Phys. Rep. \textbf{336}, 1 (2000).
\bibitem{note}There are indications \cite{prepare} that the ansatz used for calculating
shot noise in the dephasing terminal underestimates shot noise in
the limit of small $\delta x$. Obtaining the proper result for the
classical model at large $\delta x$ requires taking into account
both the anticorrelations of inputs and exchange scattering effects
at the second beam splitter (both of which reduce shot noise at intermediate
$T_{B}$).\bibitem{prepare}F. Marquardt and C. Bruder, in preparation.
\end{thebibliography}
\end{document}